\documentclass[preprint,showpacs,preprintnumbers,amsmath,amssymb]{revtex4}

\usepackage{graphicx}
\usepackage{dcolumn}
\usepackage{bm}

\begin{document}

\preprint{}

\title{Reply to Comment on "Superfluid stability in the BEC-BCS crossover"
  by Sheehy and Radzihovsky}

\author{C.-H. Pao}
\author{S.-T. Wu}
\affiliation{%
Department of Physics, National Chung Cheng University, Chiayi
621, Taiwan
}%

\author{S.-K. Yip}
\affiliation{ Institute of Physics, Academia
Sinica, Nankang, Taipei 115, Taiwan}%

\date{\today}

\begin{abstract}
The reason behind the discrepancy between the phase
diagrams of our earlier work \cite{UNpd} and 
the comment of Sheehy and Radzihovsky  \cite{SRcom}
is discussed.
We show that, in contrast to what is claimed in \cite{SRcom},
the requirement of positive susceptibility
is sufficient to rule out states that are local maximum
of the free energy (as a function of the order parameter $\Delta$).

\end{abstract}

\maketitle

It is now widely accepted that, for an attractive s-wave interaction,
an equally populated Fermi gas at zero temperature smoothly
crosses over from the BCS to the BEC regime.
In our recent paper \cite{UNpd}, we considered a two
component Fermi gas with unequal populations
under a wide Feshbach resonance.
 We showed that the uniform state must become
unstable at some intermediate coupling strength since it has either negative superfluid density or a negative susceptibility. 
Therefore we demonstrated that 
the smooth crossover known for the equal population
case is destroyed (independent of the {\it ansatz} 
what actually replaces the unstable states).
 A phase diagram was then constructed by indicating
where the uniform state was found to be unstable
(reproduced in Fig 1 here as the region between the dotted lines).

  In a recent preprint, Sheehy and Radzihovsky \cite{SRcom}
reinvestigated this phase diagram by a different method,
and found discrepancy with our earlier results \cite{UNpd}.
(see also \cite{Gu,Parish,Chien})
In particular, they found that the instability region 
occupies a larger area than ours on the BEC side of the 
the phase diagram. 
They suggest that the susceptibility criterion 
\cite{footnote} we used is not
sufficient to rule out unstable states that are actually
relative maximum of the free energy $\Omega$ with respect to
the pairing potential $\Delta$.

  We here would like to clarify the reason causing the 
discrepancy between the above two works \cite{UNpd,SRcom}.
We now believe that our phase diagram in \cite{UNpd} is incorrect. 
However, the reason was in fact due to the inaccuracies of
the numerical method we used there.  
Further, we argue that
 the susceptibility criterion is able to rule
out states that corresponds to relative maxima of $\Omega$
(such as those depicted in Fig 2 of \cite{SRcom}),
in contrast to what is claimed in \cite{SRcom}.

First we comment on the method we used in \cite{UNpd}.
A stable state must have all eigenvalues of the susceptibility
matrix $\partial n_{\sigma} / \partial \mu_{\sigma'}$ positive
(here $\sigma$ and $\sigma' = \uparrow, \downarrow$ for the two species,
and $\mu_{\uparrow, (\downarrow)} = \mu \pm h$
are their chemical potentials and $n_\sigma$ are the densities).
 In \cite{UNpd},
we solved, for the uniform states, the dimensionless 
average chemical potential 
$\tilde \mu \equiv \mu / \epsilon_F$ and
potential difference  $\tilde h \equiv h / \epsilon_F$
as a function of the dimensionless population difference
$\tilde n_d \equiv n_d / n$ and coupling constant $g \equiv 1 / k_F a$.
($n = n_\uparrow + n_\downarrow$, $n_d = n_\uparrow - n_\downarrow$).
 The {\em inverse} susceptibility matrix
$\partial \mu_{\sigma} / \partial n_{\sigma'}$
can be expressed in terms of the 
above obtained functions \cite{relations}.
It can further be shown \cite{relations} that a necessary condition
that follows is that 
$\left( \partial \tilde h / \partial \tilde n_d \right)_g$
must be positive for stability, {\it i.e.}, that the slope of
 $\tilde h$ versus $\tilde n_d$ must be
positive at fixed scaled coupling constant $g$.
Since the functions $\tilde \mu (\tilde n_d, g)$,
$\tilde h (\tilde n_d, g)$ were already available 
(see, e.g., Fig 1 of \cite{UNpd}),
we chose to evaluate this matrix and its eigenvalues numerically from
these data, and constructed the phase diagram in \cite{UNpd}.
Unfortunately, as we only found out later,
the numerical accuracies required to carry out this method is very high, 
and we erroneously concluded in \cite{UNpd} that 
one of the eigenvalue changed sign at the same position
as where $\partial \tilde h / \partial \tilde n_d$ changed sign
(footnote [13] of \cite{UNpd}).
We have now re-evaluated the positions where (one of) the
eigenvalue of this matrix changes sign more accurately, and
the result is as shown as large circles in Fig 1 in this reply.  As far
as we can tell,  this criterion yields the
same line (dot-dashed) as where $\partial^2 \Omega / \partial \Delta^2$
changes sign \cite{dOdD}.  Thus on the left of this line,
we find that our stable state has a positive susceptibility matrix
as well as corresponds to a relative minimum of the free energy $\Omega$,
whereas on the right the unstable state identified
has a susceptibility matrix with negative eigenvalue
as well as corresponds to a relative maximum of 
the free energy $\Omega$.  Therefore, the susceptibility
criterion is able to identify states that are 
relative maximum of the free energy and concludes that
they are unstable, in contrast to what was claimed in \cite{SRcom}.

  The most transparent way to see the last statement is
to consider the expression \cite{He,derivation}
\begin{equation}
\left( \frac{ \partial n_{\sigma}}{\partial \mu_{\sigma'}} \right)
=
\left( \frac{ \partial n_{\sigma}}{\partial \mu_{\sigma'}} \right)_{\Delta}
+
\frac
{ 
\left( \frac{\partial n_\sigma}{\partial \Delta} 
   \right)_{\mu_\uparrow,\mu_\downarrow} 
\left( \frac{\partial n_{\sigma'}}{\partial \Delta} 
   \right)_{\mu_\uparrow,\mu_\downarrow} 
}
{ \left(
\frac{\partial^2 \Omega}{\partial \Delta^2}
  \right)_ {\mu_\uparrow,\mu_\downarrow}
}
\label{matrix}
\end{equation}
Note that the matrix
$\left( \frac{ \partial n_{\sigma}}{\partial \mu_{\sigma'}} \right)_{\Delta}$
has eigenvalues that are positive and finite (e.g. \cite{SRcom}),
and the matrix  
$\left( \frac{\partial n_\sigma}{\partial \Delta} 
   \right)_{\mu_\uparrow,\mu_\downarrow} 
\left( \frac{\partial n_{\sigma'}}{\partial \Delta} 
   \right)_{\mu_\uparrow,\mu_\downarrow}$ has
non-negative eigenvalues.
Consider now a stable state on the BEC side at large $g$.
This state has positive susceptibilities and is a 
free energy minimum, with $\partial^2 \Omega / \partial \Delta^2 > 0$.
Consider now decreasing $g$, thus moving
towards resonance.  At the point where this state is
no longer a free energy relative minimum but just
becomes a relative maximum, 
$\partial^2 \Omega / \partial \Delta^2 < 0$ and small.
Thus the second term in eq (\ref{matrix}),
and therefore the susceptibility matrix itself, necessarily
has a large and negative eigenvalue at this point.
 Therefore, as we claimed
above, the susceptibility criterion is always able
 to identify the situation where,
as coupling constant changes, a relative minimum
becomes a relative maximum. (see also \cite{He,Chien})

  We agree however with Sheehy and Radzihosky \cite{SRcom} that
our susceptibility criterion may not always protect us
from instability caused by a first order transition.
{\em But this can occur only when a new free energy minima
arises at some other $\Delta$, rather than
 having the original minimum turning into
a relative maximum}. 
In general, our criterion can fail when one can come up
with a better {\it ansatz} for the state (at the same 
$n$ and $n_d$)
than the one originally investigated,
or there is a physical instability not contained in
the susceptibility being evaluated.
 This happens in particular
on the BCS side of the phase diagram \cite{BEC}.
  To the right of the
dotted line on the BCS side, the normal
state is a free energy relative minimum.
However, as suggested by the 
negative susceptibility matrix and  superfluid density 
(for the latter, except near $\tilde n_d \approx 1$)
of the uniform state to the left of this dotted line \cite{UNpd},
a phase separated state and a state with finite 
pairing momentum (FFLO state) are potentially more
stable than the normal state 
in a region near and to the right of this dotted line.
One can straightforwardly evaluate these instability
lines
corresponding to these two {\it ansatz}.   
For phase separation,
one finds the point where the free energy of the normal state
becomes higher (when $g$ is increased, {\i.e.}, moving
from right to left in Fig 1)
than that of the completely paired superfluid state
(at the same chemical potentials $\mu$ and $h$ of the normal state)
We evaluated the latter free energy by an 
integration over coupling constant \cite{Pao06}.
The FFLO instability line of the normal state
can be found by solving the Cooper
problem at finite pair momentum $q$, again
at the chemical potentials of the corresponding normal state.
These lines are also shown in Fig 1 here \cite{talks}.
The diagram constructed from these transition lines
then agree with \cite{SRcom} (see however the remark \cite{BEC}).

In conclusion, we have resolved 
the discrepancy between the phase diagrams
in \cite{UNpd} and \cite{SRcom}.  The reason
for the earlier disagreement was clarified.

While we were finishing the present manuscript,
we noticed another preprint \cite{Chen2} discussing
the same topic as here.  

We benefitted greatly from a private communication
with Lianyi He and Pengfei Zhuang, whom
we gratefully acknowledge.
This research was supported by the NSC of
Taiwan under grant numbers NSC94-2112-M-194-001 (CHP),
NSC94-2112-M-194-008 (STW) and NSC94-2112-M-001-002 (SKY).

\vspace{15pt}
\begin{figure}[tbh]
\begin{center}
\includegraphics[width=3in]{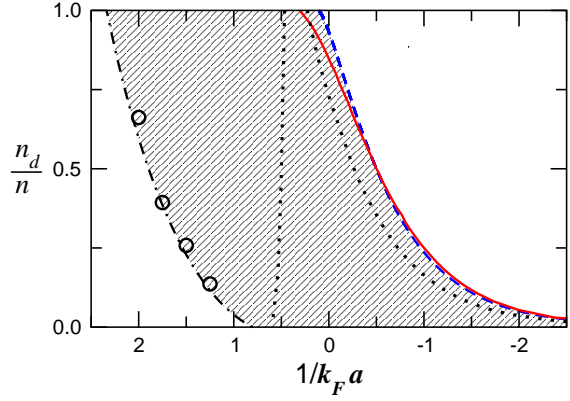}
\end{center}
 \caption{(color online) The phase diagram,
with the region where the uniform state being  
unstable shaded.  The stable region to the BCS (right) side
corresponds to the normal state, whereas the one
on the BEC  (left) side corresponds to a gapless superfluid.
 The boundaries
for the unstable region reported in \cite{UNpd}
are indicated by the dotted lines.
On the BEC side, the dot-dashed line represents 
where $\partial^2 \Omega/\partial \Delta^2$ changes sign.
Circles are where an eigenvalue of the susceptibility
matrix changes sign.   
On the BCS side, the dashed (blue) line represents the location where the
normal state becomes unstable towards phase separation.
The full (red) line represents 
the finite $q$ Cooperon instability of the normal state.}
 \label{fig:nd}
 \vspace{-5pt}
\end{figure}


\begin{thebibliography}{99}

\bibitem{UNpd} C.-H. Pao, S.-T. Wu and S.-K. Yip,
   Phys. Rev. B {\bf 73}. 132506 (2006)

\bibitem{SRcom} D. E. Sheehy and L. Radzihovsky,
  cond-mat/0608172; see also cond-mat/0607803

\bibitem{footnote} In this reply, we shall only mainly discuss
the susceptibility since the requirement of
positive superfluid density turns out to be weaker
and so only the susceptibility condition is
the subject of controversy with ref \cite{SRcom}


\bibitem{Gu} Z.-C. Gu, G. Warner and F. Zhou,
  cond-mat/0603091 

\bibitem{Chien} C.-C. Chien, Q. Chen, Y. He and K. Levin,
  cond-mat/0605039

\bibitem{Parish} M. M. Parish, F. M. Marchetti,
 A. Lamacraft and B. D. Simons,  cond-mat/0605744


\bibitem{relations} 
The inverse susceptibility matrix has elements
$ B_{\sigma \sigma'} \equiv \left(
\frac{\partial \mu_{\sigma} }{ \partial n_{\sigma'}} \right)_a
  = \left( \frac{\epsilon_F (n)}{n} \right)
  \left[ \frac{2}{3} \tilde \mu_{\sigma} -
   \frac{g}{3} \left( \frac{\partial \tilde \mu_{\sigma}}{\partial g}
      \right)_{\tilde n_d}
   + ( s - \tilde n_d) 
   \left( \frac{\partial \tilde \mu_{\sigma}}{\partial \tilde n_d}
      \right)_g \right] $
where $\tilde \mu_{\uparrow, \downarrow} = \tilde \mu \pm \tilde h$,
$s \equiv \pm 1$ for $\sigma' = \uparrow, \downarrow$,
and we have used the fact that in three dimensions,
$\epsilon_F \propto n^{2/3}$ and $k_F \propto n^{1/3}$.
The condition that all eigenvalues of $B_{\sigma \sigma'}$
are positive requires that its expectation value with respect
to the vector $(1,-1)$ be positive, and hence
$B_{\uparrow \uparrow} + B_{\downarrow \downarrow}
  - B_{\uparrow \downarrow} - B_{\downarrow \uparrow} > 0$.
Using the above expression for $B_{\sigma \sigma'}$,
we get the necessary condition
 $ \left( \frac{\partial \tilde h}{\partial \tilde n_d} 
   \right)_g > 0$.

\bibitem{dOdD} Starting from the expectation
value of the mean-field Hamiltonian in the 
gapless superfluid state, we get
$\partial^2 \Omega / \partial \Delta^2 =
\Delta^2 \sum_{\vec k} \left[
\frac{1 - f(E_k-h)}{E_k^3} + \frac{f'(E_k-h)}{E_k^2} \right]$
where $E_k$'s are the quasiparticle energies. 
(see also \cite{Chien,He})

\bibitem{He}  L. He and P. Zhuang, private communications

\bibitem{derivation} To see eq (\ref{matrix}), we note that
$\left( \frac{\partial n_{\sigma'}}{\partial \Delta} \right)
  = - \left( \frac{\partial^2 \Omega}
    {\partial \mu_{\sigma'} \partial \Delta} \right)$ and
$\frac{ \partial^2 \Omega}{ \partial \Delta^2}
   \frac{\partial \Delta}{\partial \mu_{\sigma'}} +
  \frac{\partial^2 \Omega}  {\partial \mu_{\sigma'} \partial \Delta}
  = 0$  (since $\frac{\partial \Omega}{\partial \Delta} = 0$).
 Hence eq (\ref{matrix}) reduces to
$\left( \frac{ \partial n_{\sigma}}{\partial \mu_{\sigma'}} \right)
=
\left( \frac{ \partial n_{\sigma}}{\partial \mu_{\sigma'}} \right)_{\Delta}
+ 
\left( \frac{\partial n_\sigma}{\partial \Delta} \right)
  \left( \frac{\partial \Delta}  {\partial \mu_{\sigma'}} \right) $.



\bibitem{BEC} In principle, this happens also on the BEC side
of the phase diagram, as a first order phase transition
also seems to occur there.   A more correct phase transition
line can be constructed using a procedure similar to 
Maxwell construction.
However, the position of the phase
transition line obtained by evaluating the susceptibility 
(or equivalently $\partial^2 \Omega / \partial \Delta^2 = 0$,
see also \cite{Chien,He}) turns out to be 
almost exactly the same as that given in \cite{SRcom} 
(see also \cite{Gu,Parish})
and thus the correction needed to the transition line on the BEC side
in Fig 1 is rather small.

\bibitem{Pao06} C.-H. Pao and S.-K. Yip,
  J. Cond. Matt. {\bf 18}, 5567 (2006)


\bibitem{talks}  The evaluation of these transition lines 
had already been reported 
at several international conferences 
by one of us (Yip) (Strong Correlations
in Ultracold Fermi Systems, Jan 15-21, 2006, Aspen;
and APS March 2006 Meeting, March 13-17, 2006, Baltimore.)

\bibitem{Chen2} Q. Chen, Y. He, C.-C. Chien and K. Levin,
  cond-mat/0608454

\end{thebibliography}
\end{document}